# EVM AS GENERIC QOS TRIGGER FOR HETEROGENEOUS WIRELESS OVERLAY NETWORKS

Rajender Kumar[1] and Brahmjit Singh[2]

[1]ECE Department, NIT Kurukshetra, INDIA, IEEE Member
rkumar@nitkkr.ac.in
[2]ECE Department, NIT Kurukshetra, INDIA, IETE Member
brahmjit@nitkkr.ac.in

**ABSTRACT**

*Fourth Generation (4G) Wireless System will integrate heterogeneous wireless overlay systems i.e. interworking of WLAN/ GSM/ CDMA/ WiMAX/ LTE/ etc with guaranteed Quality of Service (QoS) and Experience (QoE).QoS(E) vary from network to network and is application sensitive. User needs an optimal mobility solution while roaming in Overlaid wireless environment i.e. user could seamlessly transfer his session/ call to a best available network bearing guaranteed Quality of Experience. And If this Seamless transfer of session is executed between two networks having different access standards then it is called Vertical Handover (VHO). Contemporary VHO decision algorithms are based on generic QoS metrics viz. SNR, bandwidth, jitter, BER and delay. In this paper, Error Vector Magnitude (EVM) is proposed to be a generic QoS trigger for VHO execution. EVM is defined as the deviation of inphase/ quadrature (I/Q) values from ideal signal states and thus provides a measure of signal quality. In 4G Interoperable environment, OFDM is the leading Modulation scheme (more prone to multi-path fading). EVM (modulation error) properly characterises the wireless link/ channel for accurate VHO decision. EVM depends on the inherent transmission impairments viz. frequency offset, phase noise, non-linear-impairment, skewness etc. for a given wireless link. Paper provides an insight to the analytical aspect of EVM & measures EVM (%) for key management subframes like association/re-association/disassociation/ probe request/response frames. EVM relation is explored for different possible NAV-Network Allocation Vectors (frame duration). Finally EVM is compared with SNR, BER and investigation concludes EVM as a promising QoS trigger for OFDM based emerging wireless standards.*

## 1. INTRODUCTION

Vertical Handover (VHO) [1] is a mechanism in which user maintains connection when switched from one Radio Access Network (RAN) technology to another RAN technology (e.g., from WLAN/H-2 to UMTS and vice versa (see fig.1). [2][3] VHO is different from conventional horizontal handover where the mobile devices move from one base station to another within the same network (RAN). In VHO, a session is seamlessly handed over to a new RAN in an interoperable region based on a criterion which evaluates the signal quality. This is called 'triggering' of VHO i.e. initialising VHO. Wireless channel estimation is a complicated process and is associated with exploration of PHY and MAC layer frames therefore it is also known as L1/L2 triggering. We may include metrics/ triggers of other layers for VHO execution and can view VHO as cross layer design problem.

In Heterogeneous Wireless Overlay Network, Wi-Fi network (IEEE 802.11 b/g) is envisaged as one of the candidate wireless network (see figure-1). In interoperable wireless environment, all networks have unique uplink/downlink set of frequencies to communicate with user. In overlaid system, user has to select best available network to perform handover ensuring good quality of experience (QoE). In this regard, appropriate channel estimation is required for each wireless link & corresponding network. Objective of the paper is to investigate how better a channel can be characterised and quantified to guarantee QoE, from user's end so that an accurate decision can be made. Based on the baseband measurement of the wireless link, a new parameter, Error Vector Magnitude (EVM) is introduced. EVM is defined as the deviation of Inphase/ Quadrature (I/Q) [4] values from ideal signal states and thus provides a measure of signal quality.

EVM can be precisely termed as modulation error. EVM is quantified in percentage (%) i.e. how much percentage of deviation in modulation is observed at receiver. EVM thus rates the quality of the signal like signal to ratio, bit error rate etc. Measuring the channel quality in terms of EVM, we can therefore predict the QoE of the corresponding network (here WLAN). Similarly we can measure EVM of other collocated wireless networks and can analyse the performance. Finally, based on EVM measurement a decision can be made to select a new target network i.e. VHO decision. Emerging Wireless standards are envisaged to use Orthogonal Frequency Division Modulation (OFDM) [5]. EVM finds its vital place in OFDM based networks because they are prone to multipath fading.

The rest of the paper is organized as follows. Section 2 covers the literature survey and contemporary challenges & issues of VHO in 4G are discussed. Section 3 provides an analytical background of EVM and its relation with BER and SNR, introduced as QoS triggers. Section 4 gives performance analysis of L1/L2 frames for OFDM based wireless networks IEEE 802.11 b/g is done in. Section 5 discusses simulated results and plots. Finally paper concludes the analysis & provides future scope of the carried work.

## 2. LITERATURE SURVEY

**Next Generation Network** will have Interoperable advanced wireless systems viz.WiFi (WLAN), WSN (Wireless Sensor Networks), WiMAX [6] (Worldwide Interoperability for Microwave Access), LTE (Long Term Evolution) & etc. [7]. 4G mobile devices will be equipped with multiple network interface cards, capable of connecting to different wireless access networks i.e. Interworking Unit [IWU]. VHO operation should provide authentication of the mobile users, incur a low control overhead, and maintain the connections such that packet loss and transfer delay are minimized. VHO decision may depend on the bandwidth available for each wireless access network, the ISP (Internet Service Provider) charge for the network connection, the power usage requirements, the current battery status of the mobile device etc.

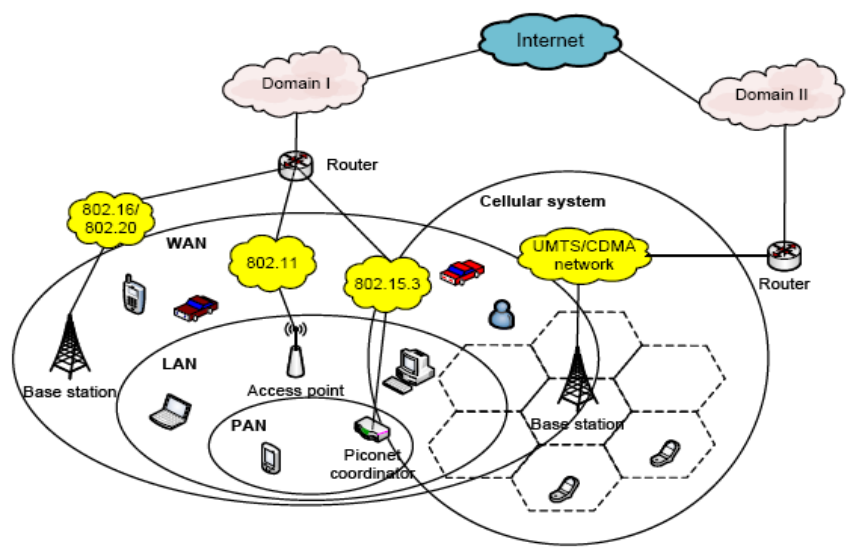

Figure 1: Heterogeneous Wireless Overlay Network

In general, the vertical handoff process can be divided into three main steps [8] namely system discovery, handoff decision, and handoff execution. There are several challenging issues on vertical handoff support [9] [10]. Limited literature is available for layer-3 VHO bearing guaranteed QoS [11] [12]. Within 3GPP (Third Generation Partnership Project) and 3GPP2 standardization groups, there are proposals describing the interconnection requirements between 3GPP systems and WLANs [13]. Within the IEEE, the 802.21 Media Independent Handover (MIH) Working Group is working towards a standard to facilitate vertical handoff between IEEE 802 technologies and 3GPP/3GPP2 networks Both QoS parameters and handoff metrics are required for vertical handoff decision. The generic QoS parameters (e.g., bandwidth, BER, SNR, delay, jitter) are specified by the applications. The information of different HO metrics is gathered during the system discovery phase.

## 2.1. VHO and generic QoS triggers

The handoff metrics and QoS parameters are categorized under different groups (e.g., bandwidth, latency, power, price, security, reliability, availability etc.). Various vertical handoff decision mechanisms have been proposed in literature based on various HO metrics and traffic classes (e.g., conversational, streaming, interactive, background) [7]. During the system discovery phase, mobile terminals equipped with multiple interfaces have to determine how many networks are available & which types of services are available in each network. The networks may also advertise the supported data rates & other QoS features for different applications. During the handoff decision phase, the mobile device determines which network it should connect to. The decision may depend on various parameters including the available bandwidth, delay, jitter, access cost, transmit power, current battery status of the mobile device, and the user's preferences. Here EVM is proposed as Generic-QoS trigger for VHO which is in good agreement to characterise a radio link and optimise VHO execution, the final phase.

## 2.2. Bit Error Rate

Bit Error Rate (BER) is a commonly used performance metric which describes the probability of error in terms of number of erroneous bits per transmitted bits. BER is a direct effect of channel noise especially for Gaussian noise channel model. For fading channels, BER performance of any communication system is worse and is related to the Gaussian noise channel performance [14].

Considering M-ary modulation with coherent detection in Gaussian noise channel and perfect recovery of the carrier frequency and phase, it can be shown that [14]

$$P_b = \frac{2\left(1-\frac{1}{L}\right)}{\log_2 L} Q\left[\sqrt{\left[\frac{3\log_2 L}{L^2-1}\right]\frac{2E_b}{N_0}}\right]$$ (2.1)

Where $L$ is the number of levels in each dimension of the $M$-ary modulation system, $E_b$ is the energy per bit and $N_0/2$ is the noise power spectral density. $Q [.]$ is the Gaussian co-error function and is given by [12]

$$Q(x) = \int_x^\alpha \frac{1}{\sqrt{2\pi}} e^{\frac{-y^2}{2}} dy$$ (2.2)

Assuming raised cosine pulses with sampling at data rate, Equation 2.1 also gives the bit error rate in terms of signal to noise ratio as

$$P_b = \frac{2\left(1-\frac{1}{L}\right)}{\log_2 L} Q\left[\sqrt{\left[\frac{3\log_2 L}{L^2-1}\right]\frac{2E_s}{N_0 \log_2 M}}\right]$$ (2.3)

Where $E_s/N_0$ is the signal to noise ratio for the *M*-ary modulation system and raised cosine pulse shaping at data rate. Equation 2.3 defines the BER performance in terms on SNR and quite often used as main tool for many adaptive systems. For diversity and MBER systems, this equation essentially means that the choice is made in favour of bit error rate.

## 3. ERROR VECTOR MAGNITUDE

### 3.1. EVM definition

Error Vector Magnitude (EVM) is initially modelled by [K.M. Ghairabeh, K.G. Gard, and M.B. Steer [15]-[18]. Measurements are performed on vector signal analyzers (VSAs), real-time analyzers which capture a time record and internally perform a Fast Fourier Transform (FFT) to enable frequency domain analysis. Signals are down converted before EVM calculations are made. Since different modulation systems viz. BPSK, 4-QAM, 16- QAM etc. have different amplitude levels, to calculated and compare EVM measurements effectively some normalization is typically carried out [19]. The normalization is derived such that the mean square amplitude of all possible symbols in the constellation of any modulation scheme is one. Thus, EVM is defined as the root-mean-square (RMS) value of the difference between a collection of measured symbols and ideal symbols. These differences are averaged over a given, typically large number of symbols and are often shown as a percent of the average power per symbols of the constellation. As such EVM can be mathematically given as

$$EVM\,rms \approx \left[ \frac{\left[\frac{1}{N}\sum_{n=1}^{N}\left|S_n - S_{0,n}\right|^2\right]}{\left[\frac{1}{N}\sum_{n=1}^{N}\left|S_{0,n}\right|^2\right]} \right] \quad (2.4)$$

Where $S_n$ is the normalized nth symbol in the stream of measured symbols, $S_{0,n}$ is the ideal normalized constellation point of the nth symbol and N is the number of unique symbols in the constellation. The expression in Equation 2.4 cannot be replaced by their unnormalized value since the normalization constant for the measured constellation and the ideal constellation are not the same. The normalization scaling factor for ideal symbols is given by [20]

$$|A| = \sqrt{\frac{1}{\frac{P_v}{T}}} = \sqrt{\frac{T}{P_v}} \quad (2.5)$$

Where $P_v$ is the total power of the measured constellation of T symbols. For RMS voltage levels of Inphase and Quadrature components, $V_I$ and $V_Q$ and for T >> N, it can be shown that $P_v$ can be expressed as

$$P_v = \sum_{t=1}^{T}\left[(V_{I,t})^2 + (V_{Q,t})^2\right](W), \quad (2.6)$$

The normalization factor for ideal case can be directly measured from *N* unique ideal constellation points and is given by

$$|A_0| = \sqrt{\frac{N}{\sum_{n=1}^{N}\left[(V_{I0,n})^2 + (V_{Q0,n})^2\right]}} \quad (2.7)$$

Hence Equation 2.4 can be further extended using normalization factors in Equations 2.5.
Modulation Error Rate (EVM): Root mean square value of error, Error$_{rms}$

$$\left[\frac{1}{N_f}\right]\sum_{i=1}^{N_f}\sqrt{\sum_{j=1}^{L_p}\left(1/52 L_p P_o\right)\left[\sum_{k=1}^{52}\left\{(A)^2+(B)^2\right\}\right]}.\qquad(2.8)$$

$A= I(i,j,k) - Io(i,j,k)$ ; $B= Q(i,j,k) - Qo(i,j,k)$

Where, $L_p$ is the length of the packet.
$N_f$ is the number of frames for the measurement.
$I_o(i,j,k,)$, $Q(i,j,k)$ denotes the ideal symbol point of the $i^{th}$ frame, $j^{th}$ OFDM symbol of the frame, $k^{th}$ subcarrier of the observed point of the complex plane.
$(I(i,j,k,), Q(i,j,k))$ denotes the observed point of the $i^{th}$ frame, $j^{th}$ OFDM symbol of the frame, $k^{th}$ subcarrier of the OFDM symbol in the complex plane. $P_o$ is the average power of the constellation. This is the definition which is now being used as the standard definition of the EVM in IEEE 802.11a – 1999*TM* [21].

### 3.2. Relationship between EVM and BER

From Equation 2.8, it is evident that EVM is essentially the normalized error magnitude between the measured constellation and the ideal constellation. For Gaussian noise model,

$\text{EVM}_{rms} \approx (1/\text{SNR})^{1/2} \approx (No/Es)^{1/2}$ \hfill (2.9)

In order to establish relationship between BER and EVM, SNR in Equation 2.10 can be expressed in terms of EVM as

$\text{SNR} \approx [1/(\text{EVM}^2)]$ \hfill (2.10)

Combining Equations 2.11 and 2.3, we can now relate the bit error rate ($P_b$) directly with the error vector magnitude as follows

$$P_b \approx \frac{2(1-(1/L))}{\log_2 L}.Q\left[\sqrt{\left[\frac{3.\log_2 L}{L^2-1}\right]}.\frac{2}{Log_2 M.EVM_{rms}^2}\right]$$

(2.11)

## 4. PERFORMANCE ANALYSIS (L1/L2 FRAMES)

The use of a layered architecture is consistent with the design methodologies employed in the current IEEE 802-based systems [22]. Together, the PHY and MAC layers, also known as layer 1 and 2, make up the services to be delivered. The MAC layer should be optimized to support a specific PHY implementation. If more than one PHY implementation is to be used, the MAC layer should be designed to have a PHY-specific layer. It is envisioned to follow a modular approach for designing such a network. There should be a clear separation of functionality in the system between the user, data, and control. The PHY and MAC layers each should have a set of well defined responsibilities that are encapsulated within the respective layers. Simulation is focused on inherent transmission impairments of wireless links, and user mobility i.e. baseband measurements of PHY and MAC layer parameters of IEEE 802.11 b/g [7].

During the HO execution phase, connections need to be re-routed from the existing network to the new network in a seamless manner. This phase also includes the authentication and authorization, and the transfer of user's context information. In MAC layer configuration, management and control frames initiates the association of user with target network. In this paper, EVM is measured for Management frames only.

### 4.1. Management Frames

Management Frames contain information for the receiving MAC management entity [7][22][23]. Management frames viz. Association request/ response, Reassociation request response, Reassociation response, Authentication,De-authentication , Probe request/ response, Beacon Frames etc. IEEE 802.11 transmitter sends an association request to 802.11 receivers or an access point to begin the association process. This frame carries information about the source such as supported data rates and the SSID of the network it wishes to associate with. After receiving the association request, the receiver or an access point considers associating with the NIC, and (if accepted) reserves memory space and establishes an association ID for the source.

Here PHY layer and MAC layers are configured for Management frames for infrastructure networks. For the performance comparison the modulation formats used is OFDM [11]. Here, Management frames are set to compute the EVM for fix data payload (10 Kb) subjected to impairments. OFDM systems are prone to multipath fading effects which degrade signal quality. Channel estimation (tracking) in OFDM systems is generally based on the use of pilot subcarriers. If pilot tracking is enabled, we can observe a significant improvement in the overall EVM performance.

EVM gives the measure of modulation error. The modulation error indicates the deviation of In phase and Quadrature phase (I/Q) values from ideal signal states and thus provides a measure of signal quality. WLAN devices operate in the 2.4 GHz and 5 GHz bands with bandwidths of 20 MHz.

Measurements are taken for EVM, by varying NAV (Network allocation Vector). NAV is Duration /ID field depending upon type of frame. This field has two meanings depending on the frame type: In power save Poll messages this is the station ID and in all other frames this is duration value in micro seconds used for the NAV calculation.

The analysis is extended to adhoc networks. IEEE802.11 frame format component.

$OH_{mgt}$ = Management Overhead bits = Preamble +PLCP Header + CRC+ MAC data [8] [24] [25].

### 4.2. Configuration used for Simulation: Parameters & attributes

- Network Type: Infrastructure
- Standard-IEEE 802.11g
- Modulation scheme-OFDM
- Data rate-06 Mbps
- Test signal – 11110000
- Channel impairment--AWGN (6db)
- Freq offset (1 KHz)
- EVM ($E_{rms}$)
- EVM pilot ($E_{pil}$)
- EVM ($E_{db}$)
- EVM pilot (Ep_db)
- Frequency error (F_err)
- Code rate (0.5)
- Frame length (IEEE std)
- Scrambler is enabled
- MAC parameters: Phase and Amplitude tracking is enabled.

Measurement performed in the paper is done on WilANTA Simulator from Seasolve's (USA). Simulator comprises of vector signal generator (VSG), spectrum analyser and vector signal analyser (VSA) [23].

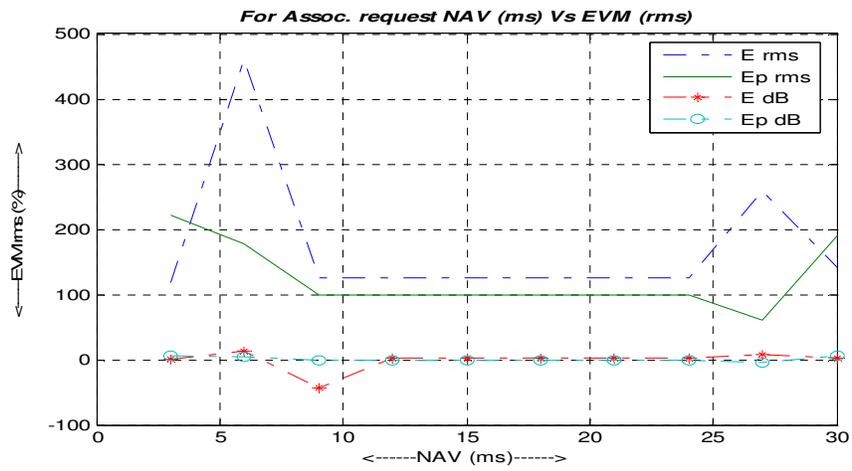

Fig 2: EVM for Association request subframes

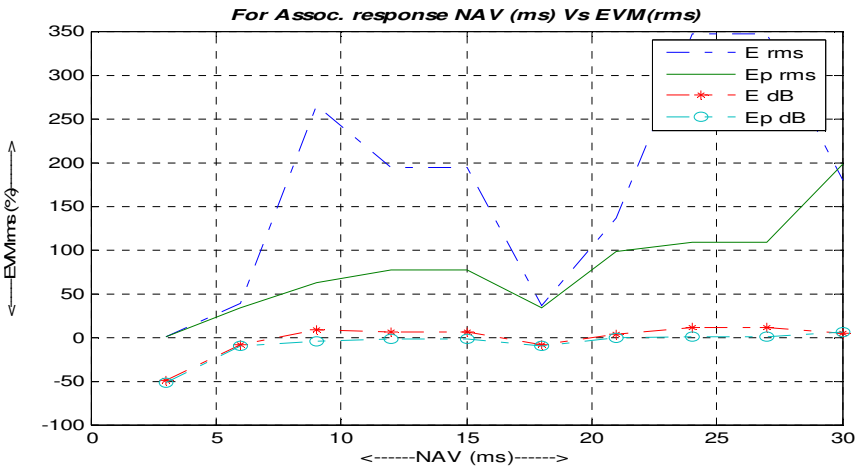

Fig 3: EVM for Association response subframes

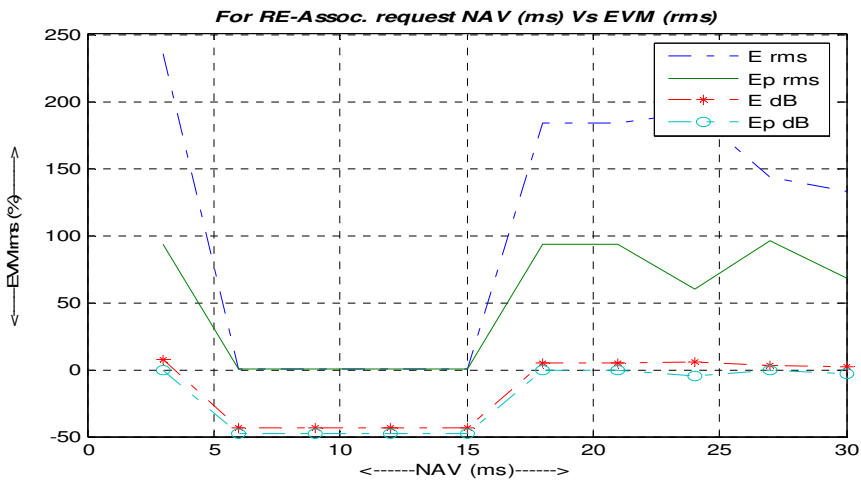

Fig 4: EVM for RE-Association request subframes

International Journal of Wireless & Mobile Networks ( IJWMN ), Vol.2, No.3, August 2010

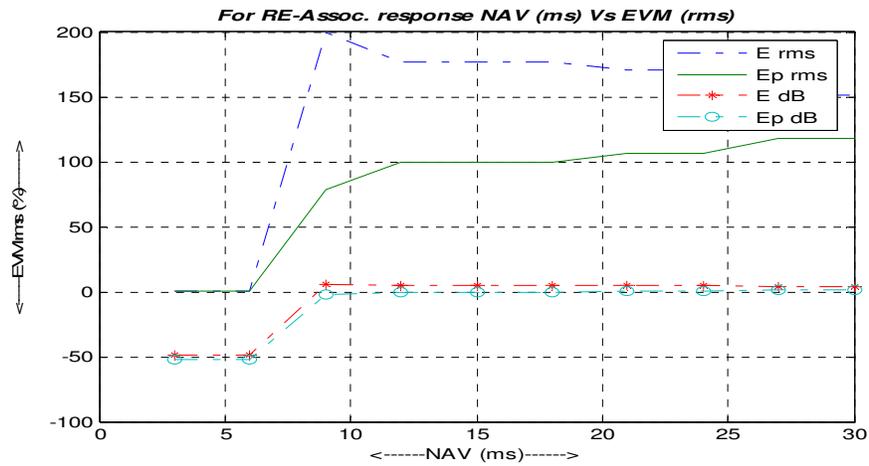

Fig 5: EVM for RE-Association response subframes

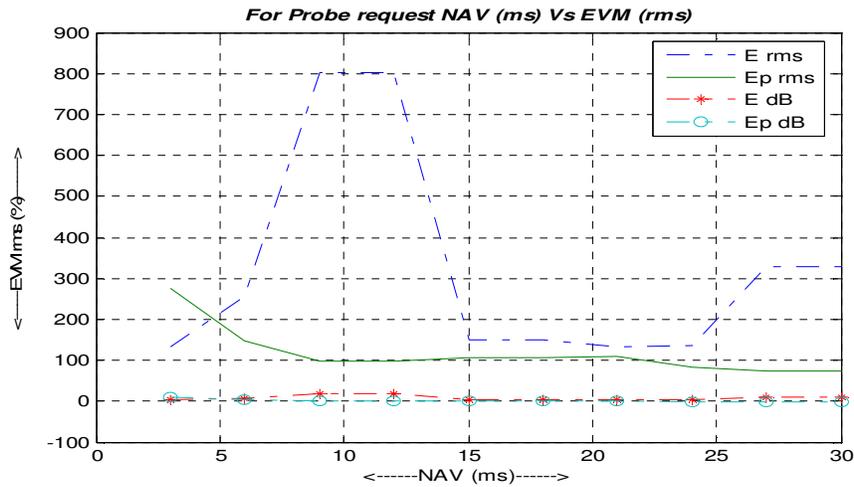

Fig 6: EVM for Probe request subframes

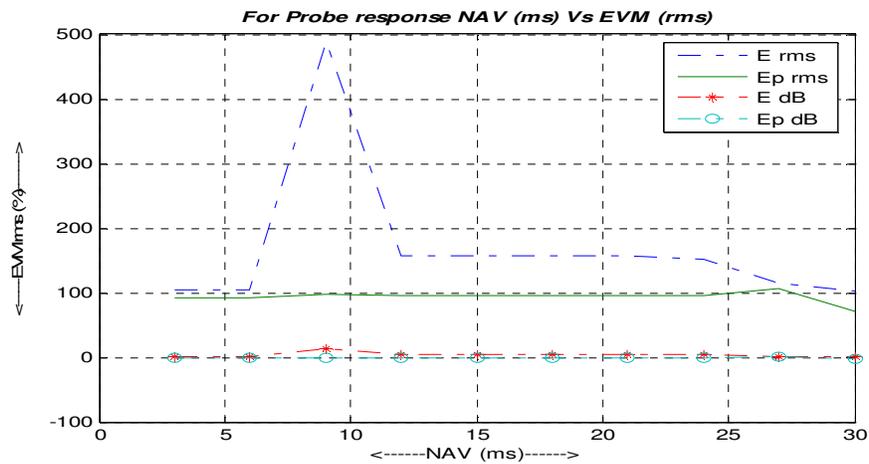

Fig 7: EVM for Probe response subframes

## 5. RESULT & DISCUSSION

**5.1.** Management frames (sub-frames) are used for the analysis where user has to associate/ disassociate/ re-associate with a new access point (AP) for VHO decision. When a mobile station asks for permission to associate with a new AP, it sends association frames. For pre-mature connection and disconnection, re-association frames are sent. Finally disassociation frames are sent to terminate a session/ call from the serving AP.

**5.2.** Control & (especially) Data-frames reception are further degraded when subjected to similar channel impairment. $E_b/N_o$ or signal power needs to be enhanced in order to overcome the impairment viz. caused due to frequency offset. Results show an abnormal rise in EVM because frames are experiencing hostile environment. EVM is observed low when $E_b/N_o$ is increased which means there is subsequent reduction in EVM at higher signal power level thereby improving reception quality.

**5.3.** From the subframes discussed in section 3 and Fig.2 to Fig.7, it is observed that there is an increase in $EVM_{dB}$ and $EVM_{rms}$ value(s) with the increase in NAV durations i.e. when NAV is greater than 15-20 millisecond, the $EVM_{rms}$ is increasing abruptly above 50% (refer Fig 3 and Fig 4 to 7). Thus, for long frame formats assigned to association/ re-association and probe subframes, receiver senses an increase in modulation error though data rate is low (6/9 mbps).

**5.4.** For management and control frames, EVM are minimized which shows performance improves when we introduce pilot sub-carrier/ signal (green solid line ---). EVM is measured for both the cases i.e. with and without pilot subcarrier. With pilot subcarrier EVM is reduced resulting in improved performance. Here performance is improved (reduced EVM) but it is not desired at the cost of an extra pilot carrier. Additional pilot carrier requirement is not an efficient approach as carrier is scarce resource. [See section 3].

**5.5.** All the subframes with varied NAV (except probe request) had EVM above 50 %. This Suggests, Frame reception can be made better if we keep low NAV value less than ~17 ms.

**5.6.** Compared to the bit error rate (BER), which gives a simple one-to-one binary decision as to whether a bit is erroneous or not, EVM is basically a measure of errors between the measured symbols and expected symbols. Due to the simplicity of comparison, BER has been a major choice to engineers, industries and researchers. BER is related to signal to noise ratio (SNR), since direct relationship exists between them. However, for BER measurement, it is obvious that signal must be demodulated first at the receiver side. This means that for every update in the adaptive algorithm, it has to receive feedback from the receiver end [8]. In 4G, OFDM is proposed to be the principle modulation scheme for emerging standards viz. IEEE 802.16e, IEEE 802.21 and LTE including WLAN [24]. OFDM systems are prone to multipath fading thus EVM finds its maximum utility i.e. it can accurately characterise the OFDM link & address the overall Network performance for VHO.

## CONCLUSION

EVM and Frequency Errors should be minimized in order to enhance the performance of data networks. Measurement results shows that EVM is a suitable substitute as compared to contemporary metric characterising a radio link viz. SNR, BER, delay, bandwidth, jitter etc. VHO decision might fall short if proper QoS metric or combination of metrics is not chosen and measured properly. Paper shows different pilot tests to propose EVM as Quality of Service parameter to be interrogated and measured during essential MAC frame transaction between MS & AP during VHO. It is essential that a given radio link under consideration is properly characterise by a metric or set metrics. Otherwise user may lead to wrong selection of (target) network. Appropriate target network selection is the key concern for seamless transfer of any ongoing session in heterogeneous wireless overlay networks. For VHO initiation, it's not always necessary to gather feedback from received end; rather EVM can give the desired performance metric before the demodulation can actually takes place. Thus novelty of the research contribution is that proposed EVM can optimise the VHO algorithm, when used as **VHO trigger**.

## FUTURE SCOPE OF WORK

In future scope of the paper, VHO utility function may be developed for the emerging heterogeneous wireless systems (Wi-Fi/ WiMAX/ LTE). Mobility management may further be expedited based on EVM trigger. Work may be extended to Layer-3 in conjunction with Layer -1 & 2 triggering of VHO and can be viewed as cross layer design problem.

## REFERENCES

[1] J. McNair and F. Zhu, "Vertical Handoffs in Fourth-generation Multinetwork Environments," IEEE Wireless Comm., vol. 11, no. 3, June 2004.

[2] IEEE, "Draft IEEE Standard for Local and Metropolitan Area Networks: Media independent Handover Services," P802.21/D00.01, July 2005.

[3] 3GPP, "Requirements on 3GPP System to Wireless Local Area Network (WLAN) interworking," TS 22.234 (v7.0.0), January 2005.

[4] Andrea Goldsmith. Wireless Communications. Cambridge University Press, Stanford University, 1$^{st}$ edition, 2005.

[5] Rui Wang, Vincent K. N. Lau, Linjun Lv, and Bin Chen Joint Cross-Layer Scheduling and Spectrum Sensing for OFDMA Cognitive Radio Systems IEEE transactions on wireless communications, pp. 2410-2416, vol. 8, no. 5, may 2009.

[6] Claudio Cicconetti, Luciano Lenzini, and Enzo Mingozzi, University of Pisa Carl Eklund, Nokia Research Center , "Quality of Service Support in IEEE 802.16 Networks", IEEE Network, 50-55. March/April 2006.

[7] Hossam Fattah and Hussein Alnuweiri, 'A Cross-Layer Design for Dynamic Resource Block Allocation in 3G Long Term Evolution System',pp.929-934, IEEE, 2009.

[8] Rajender Kumar, Brahmjit Singh, on 'Performance characteristic measurements of Management frames in IEEE 802.11x for VHO decision' International Conference on Wireless Mobile and Multimedia System, WMMIC08-Mumbai, Paper ID-48, pp. Pp.210-214, 11-12$^{TH}$ Jan 2008, IET (IEE) Sponsored.

[9] Victoria Fineberg, A Practical Architecture for Implementing End-to-End QoS in an IP Network, IEEE Communications Magazine • ,pp. 122-130 January 2002.

[10] Method of Identifying a radio link, Inventor: Oliver, Stuttgart (DE), US 7,616, 928 B2, date of patent Nov. 10, 2009.

[11] Patent no. US 7,613,457 B2; Inventors: Xia Gao, Campbell, CA (US); Gang Wu, Cupertino, CA(US)systems and method for supporting Quality of Service in vertical handovers between heterogeneous networks, , date of patent—Nov. 3, 2009.

[12] Patent Pub. No.: US 2009/ 0285176, Inventors: Halhong Zheng, Coppell, TX (US); Shashikant Maheshwari, Irving, TX (US), Basavaraj Patil, Coppell, TX (US); Srinivas Sreemanthula, Flower Mound, TX (US); Srinivas Sreemanthula, Flower Mound, TX (US), 'Framework for Internetworking between WMAN & WLAN networks' , dated- Nov. 19, 2009.

[13] 3GPP2, "3GPP2-WLAN Interworking," S.R0087-0 (v1.0), July 2004.

[14] Lajos Hanzo, William Webb, and Thomas Keller. Single- and Multi-Carrier Quadrature Amplitude Modulation. Wiley, Chichester, 2nd edition, 2000.

[15] Gharaibeh, K.M., Gard, K.G., and Steer, M.B.: 'In-band distortion of multisines', IEEE Trans. Microwave Theory Tech., 8, pp. 3227–3236, 2006.

[16] Apostolos Georgiadis, Gain, Phase Imbalance, and Phase Noise Effects on Error Vector Magnitude, IEEETransactions on Vehicular Technology, Vol. 53, No. 2, Mar 2004.

[17] Gharaibeh K.M., and Steer, M.B.: 'Modeling distortion in multi-channel communication systems', IEEE Trans. Microw. Theory Tech., , 53, (5), pp. 1682–1692, 2005.

[18] S. Forestier, P. Bouysse, R. Quere, A. Mallet, J.-M. Nebus, and L. Lapierre, Joint optimization of the power-added efficiency and the error-vector measurement of 20-GHz pHEMT amplifier through a new dynamic bias-control method, IEEE Trans. Microwave Theory and Tech, vol. 52, no.4, pp. 1132-1140, Apr. 2004.

[19] K.M. Ghairabeh, K.G. Gard, and M.B. Steer. "Accurate Estimation of Digital Communication System Metrics - SNR, EVM and _ in a Nonlinear Amplifier Environment". IEEE Transactions on Communications, pages pp.734–739, Sept. 2005.

[20] Lavrador, P.M., de Carvalho, N.B., and Pedro, J.C.: 'Evaluation of signal-to-noise and distortion ratio degradation in nonlinear systems', IEEE Trans. Microw. Theory Tech., 2004, 52, (3), pp. 813–822

[21] IEEE, IEEE Standard 802.11b-1999. IEEE Standard for Wireless LAN Medium Access Control (MAC) and Physical Layer (PHY) Specifications: High Speed Physical Layer Extension in the 2.4GHz Band.

[22] IEEE, IEEE Standard 802.11a-1999. IEEE Standard for Wireless LAN Medium Access Control (MAC) and Physical Layer (PHY) Specifications: High Speed Physical Layer in the 5GHz Band.

[23] WiLANTA simulator (VSG/VSA), Seasolve (USA).

[24] Method of Identifying a radio link, Inventor: Oliver, Stuttgart (DE), US 7,616, 928 B2, date of patent Nov. 10, 2009.

[25] Rui Wang, Vincent K. N. Lau, Linjun Lv, and Bin Chen Joint Cross-Layer Scheduling and Spectrum Sensing for OFDMA Cognitive Radio Systems IEEE transactions on wireless communications, pp. 2410-2416, vol. 8, no. 5, may 2009.

[26] Jen--Chu LiuChu LiuMNET Lab, CS, NTHUMNET NTHU , "Quality of Service Provisioning in WiMAX Networks: Chances and Challenges".Aug 2007

[27] Claudio Cicconetti, Luciano Lenzini, and Enzo Mingozzi, University of Pisa Carl Eklund, Nokia Research Center , "Quality of Service Support in IEEE 802.16 Networks", IEEE Network, 50-55. March/April 2006

[28] I. Akyildiz, J. Xie, and S. Mohanty, "A Survey of Mobility Management in Next-Generation All-IP-Based Wireless Systems," IEEE Wireless Communications, vol. 11, no. 4, August 2004.

[29] RFC 791, RFC 2474.

# Author's Profile

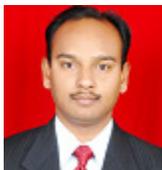  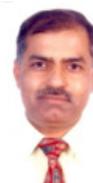

**Rajender Kumar, Asstt. Professor**     **Dr. Brahmjit Singh, Professor**

### Corresponding Author Biography: Rajender Kumar

**Rajender Kumar is working** as Asstt. Professor in Electronics and Communication Engineering Department at NIT Kurukshetra. He is pursuing PhD in Vertical Handovers (Heterogeneous Wireless Networks). He did B.Tech in Electronics and Communication Engineering (ECE) from Institute of Engineering Technology, Kanpur in 2001. He received M.Tech (Hons.) in ECE from National Institute of Technology, Kurukshetra, (INDIA) in 2004. He worked in Indian Telephone Industries as NSS (switch) Engineer at Pune, in GSM (Mobile) Division. He has 6+ years of teaching/ industrial and research experience. He teaches Telecommunication networks, Statistical modeling, Mathematical models for internet and web in M.Tech program. He takes Computer communication networks and Signals & system in B.Tech program. He has 14 research paper publications in National and International Conferences and Journal. His field of Interest includes Vertical handovers, Advanced wireless systems, Cognitive radio, NGWN challenges & Cross layer design issues. He is reviewer and consultant of leading publisher and government organization like Tata Mc-Graw Hill publishers & KDB respectively. Rajender Kumar received 'All India Young Scientist Award' in Silver Jubilee function held at Bhopal in November, 2007.

### Co-Author Biography: Brahmjit Singh

**Brahmjit Singh** received B.E. degree in Electronics Engineering from Malaviya National Institute of Technology, Jaipur in 1988, M.E. degree in Electronics and Communication Engineering from Indian Institute of Technology, Roorkee in 1995 and PhD degree from Guru Gobind Singh Indraprastha University, Delhi in 2005 (India). He started his career as a lecturer at Bundelkhand Institute of Engineering and Technology, Jhansi (INDIA). Currently, he is Dean & Professor in School of ICT at Gautam Buddha University, Noida (INDIA). He was Ex-head of CCN & ECE Department at National Institute of Technology, Kurukshetra (INDIA). He teaches post-graduate and graduate level courses on Wireless communication and CDMA systems. His research interests include Mobility management in cellular / wireless networks, Planning, Designing and optimization of Cellular networks and Wireless network security. He has published ~ 45 research papers in International/ National journals and Conferences. Dr. Brahmjit Singh received the Best Research Paper Award from 'The Institution of Engineers (India)' in 2006.